\documentclass{Pos}

\usepackage{amssymb}
\usepackage{fontenc}
\usepackage{times}
\usepackage{mathptmx}
\usepackage{graphicx}
\usepackage{amsmath}

\title{Nucleon-to-pion transition distribution amplitudes and backward electroproduction of pions}

\ShortTitle{$\pi N$ TDAs and backward electroproduction of pions}

\author{B. Pire\\
        CPhT, \'{E}cole Polytechnique, CNRS,  91128 Palaiseau, France \\
}

\author{\speaker{K.~Semenov-Tian-Shansky}
\\
       IFPA, d\'{e}partement AGO,  Universit\'{e} de  Li\`{e}ge, 4000 Li\`{e}ge,  Belgium\\
        E-mail: \email{ksemenov@ulg.ac.be}}

\author{L.~Szymanowski \\
        National Center for Nuclear Research (NCBJ), Warsaw, Poland\\
}

\abstract{

Baryon to meson
transition distribution amplitudes (TDAs), non-diagonal matrix
elements of the nonlocal three quark operator between a nucleon and
a meson state, extend the concept of generalized parton distributions.
These non-perturbative objects which encode the information on three quark
correlations inside the nucleon may be accessed experimentally in backward meson
electroproduction reactions.
We suggest a general framework for  modelling nucleon to pion ($\pi N$) TDAs
employing the spectral representation for
$\pi N$
TDAs in terms of quadruple distributions.
The factorized Ansatz for quadruple distributions with input
from the soft-pion theorem for
$\pi N$ TDAs is proposed.  It is to be complemented with a
$D$-term like contribution from  the nucleon exchange in the cross channel.
We present our estimates of  the unpolarized cross
section  and of the transverse target single spin asymmetry for  backward pion electroproduction
within the QCD collinear factorization approach in which the non-perturbative
part of the amplitude involves $\pi N$ TDAs.
The cross section is sizable enough to be studied in high luminosity experiments
such as J-lab@12GeV and EIC.

}

\FullConference{Sixth International Conference on Quarks and Nuclear Physics,\\
		April 16-20, 2012\\
		Ecole Polytechnique, Palaiseau, Paris}

\begin{document}

\section{Introduction}
\begin{figure}
\begin{center}
\includegraphics[width=.3\textwidth]{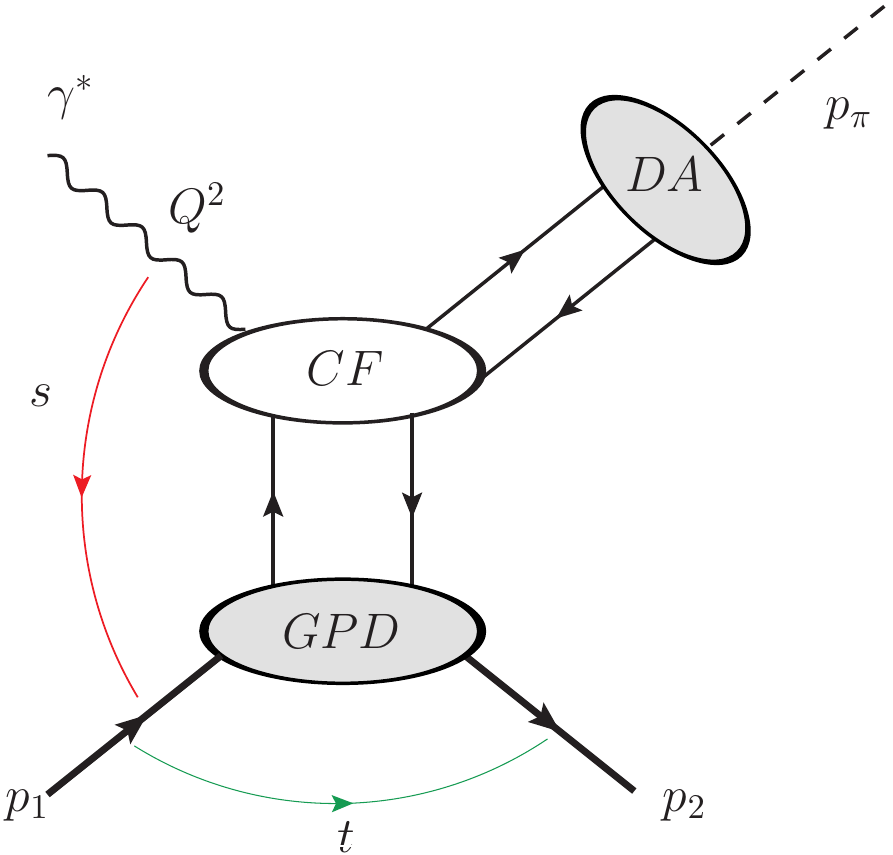} \ \ \ \ \ \ \ \ \
\includegraphics[width=.3\textwidth]{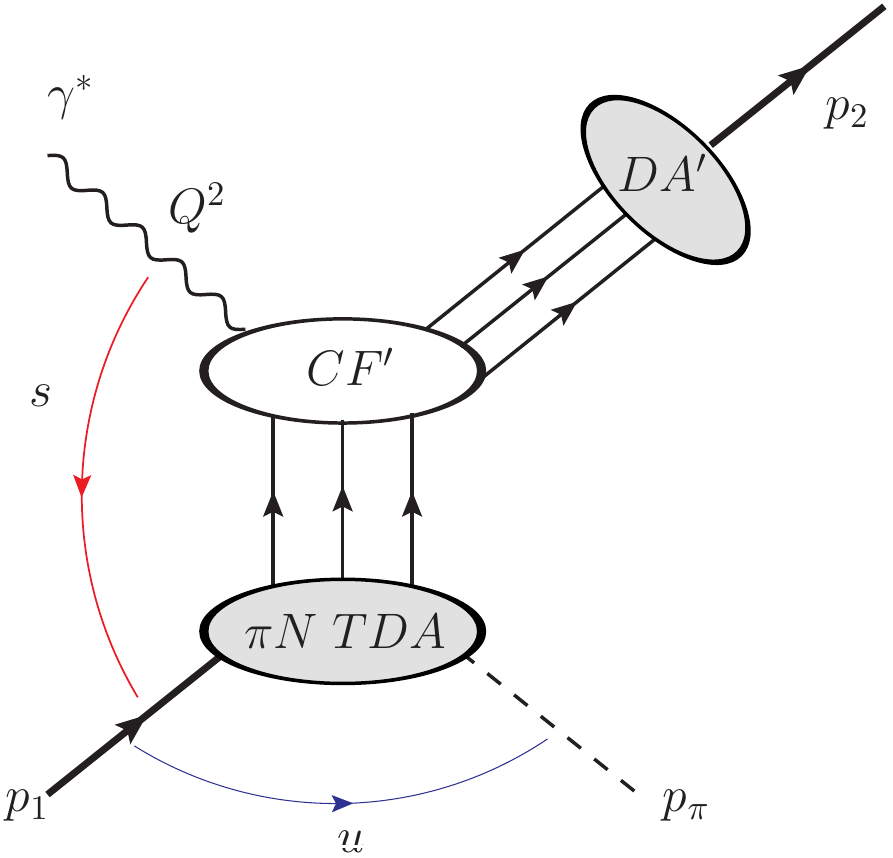}
\caption{Collinear factorization for hard production of pions off nucleon in the
conventional hard meson production (HMP) kinematics ( {\bf left}) versus the collinear factorization
in the backward kinematics regime ( {\bf right}). DA (DA') denote pion (nucleon) distribution amplitude;
CF (CF') are coefficient functions computable in perturbative QCD. }
\label{fig1}
\end{center}
\end{figure}

The backward kinematical regime for the pion electroproduction off nucleons
 (\ref{reaction})
\begin{equation}
e(k_1)+ N(p_1) \rightarrow \big( \gamma^*(q) + N(p_1) \big) +e(k_2)
  \rightarrow  e (k_2)+ \pi(p_\pi) + N'(p_2).
\label{reaction}
\end{equation}
provides experimental access to nucleon to pion transition distribution amplitudes ($\pi N$ TDAs).
Backward kinematics implies that $Q^2=-q^2$ and $s=(p_1+q)^2$ are large, $x_{B}=\frac{Q^2}{2 p_1 \cdot q}$
and skewness variable are kept fixed 
(skewness is defined with respect to the $u$-channel momentum transfer
$\Delta = p_\pi-p_1$:
$\xi=-\frac{\Delta \cdot n}{(p_\pi-p_1) \cdot n}$, where $n$ is the conventional light cone vector occurring in the Sudakov decomposition of the momenta);
the  $u$-channel momentum transfer squared 
$u \equiv \Delta^2$ is supposed to be small ($|u| \sim 0$ corresponds to a pion
produced in a near backward direction in $\gamma^* N$ center-of-mass frame).
$\pi N$ TDAs first considered in \cite{Frankfurt:1999fp}
arise within the collinear factorization approach for the reaction (\ref{reaction})
in the backward kinematics (see right panel of Fig~\ref{fig1}).
It's worth to specially emphasize that backward kinematics is complementary to the conventional
generalized Bjorken limit
(large $Q^2$ and $s$; fixed $x_{Bj}$ and skewness defined with respect to the $t$-channel momentum transfer
$\xi=-\frac{(p_2-p_1) \cdot n}{(p_1-p_1) \cdot n}$
and small $t$-channel momentum transfer squared $|t| \sim 0$)
in which the factorized description \cite{Collins:1996fb}  in terms of generalized parton distributions (GPDs)
applies to the reaction (\ref{reaction}).

$\pi N$ TDAs may be seen as further development of the GPD concept. They are defined through
the $\pi N$ matrix element  of the three-local quark operator on the light-cone
\cite{Efremov:1978rn},
\cite{Lepage:1980}:
\begin{equation}
\hat{O}^{\alpha \beta \gamma}_{\rho \tau \chi}( \lambda_1 n,\, \lambda_2 n, \, \lambda_3 n) =
\Psi^\alpha_\rho(\lambda_1 n)
\Psi^\beta_\tau(\lambda_2 n)
\Psi^\gamma_\chi (\lambda_3 n),
\label{oper}
\end{equation}
were
$\alpha$, $\beta$, $\gamma$
stand for quark flavor indices and
$\rho$, $\tau$, $\chi$
denote the Dirac spinor indices; antisymmetrization in color is implied;  and the gauge links
are omitted in  the light-like gauge
$A \cdot n=0$.

The extensive studies of the properties and physical interpretation of $\pi N$ TDAs are presented in
Refs. \cite{Pire:2005ax,Lansberg:2007ec,Lansberg:2007se,Pire:2010if,Pire:2011xv,Lansberg:2011aa}.
Conceptually, $\pi N$ TDAs share common features both with GPDs and nucleon distribution amplitude.
Indeed, the crossing transformation relates $\pi N$ TDAs with $\pi N$ generalized distribution amplitudes
(GDAs), defined as the matrix element of the same light cone operator between $\pi N$ state and the vacuum.
In their turn, $\pi N$ GDAs reduce to combinations of the usual nucleon DA in the soft pion limit.
On the other hand, similarly to GPDs
\cite{Burkardt:2000za},
a comprehensible physical picture  may be obtained by switching to the
impact parameter space. Baryon to meson TDAs are supposed to encode new informations on the
hadron structure in the transverse plane.
There are hints
\cite{Strikman:2009bd}
that $\pi N$ TDAs
may be used as a tool to perform the femto-photography \cite{Ralston:2001xs}
of nucleon's pion cloud.

\section{$\pi N$ TDAs: theoretical constrains and modelling }

Below we  summarize the fundamental requirements for $\pi N$ TDAs which follow from the symmetries of QCD
established in Refs.
\cite{Pire:2010if,Pire:2011xv,Lansberg:2011aa}.

\begin{itemize}
\item  For given flavor contents spin decomposition of the leading twist-$3$
$\pi N$
TDA involve eight invariant functions
$V_{1,2}^{\pi N}$, $A_{1,2}^{\pi N}$, $T_{1,2,3,4}^{\pi N}$
each depending on the  longitudinal momentum fractions
$x_i$ ($\sum_{i=1}^3 x_i=2\xi$),
skewness parameter
$\xi$
and the $u$-channel momentum transfer squared
$\Delta^2 \equiv (p_\pi-p_1)^2$
as well as on the factorization scale $\mu^2$.
\item  Isotopic and permutation symmetries
 reduce the number of independent leading twist $\pi N$ TDAs
to just eight:
four in both  the isospin-$\frac{1}{2}$
and the isospin-$\frac{3}{2}$.

\item The support of $\pi N$ TDAs in three longitudinal momentum fractions
$x_i$
is given by the intersection of  the stripes
$-1+\xi \le x_i \le 1+\xi$ ($\sum_{i=1}^3 x_i=2\xi$).
One can distinguish the Efremov-Radyushkin-Brodsky-Lepage-like (ERBL-like) domain,
in which all
$x_i$
are positive and two type of Dokshitzer-Gribov-Lipatov-Altarelli-Parisi-like (DGLAP-like) domains, in which
one or two
$x_i$
turn negative.

\item The evolution properties of
$\pi N$
TDAs are described by the appropriate generalization
\cite{Pire:2005ax}
of the  ERBL/ DGLAP evolution equations specific for the domain in $x_i$.

\item Underlying Lorentz symmetry results in the polynomiality property for the Mellin moments of
$\pi N$
TDAs
in the longitudinal momentum fractions
$x_i$.
Similarly to the GPD case, the
$(n_1,\,n_2,\,n_3)$-th ($n_1+n_2+n_3 \equiv N$)
Mellin moments of nucleon to meson TDAs  in
$x_1$, $x_2$, $x_3$
are polynomials of powers
$N$ or $N+1$
in the skewness variable
$\xi$.

\item Crossing transformation relates
$\pi N$
TDAs to
$\pi N$
GDAs, defined by the matrix element of the same operator
(\ref{oper})
between the
$\pi N$
state and  the vacuum. Therefore the soft pion theorem
\cite{KSemenov_Pobylitsa:2001cz}
for $\pi N$ GDAs
\cite{KSemenov_Braun}
expresses
$\pi N$
TDAs at the soft pion threshold
$\xi=1$, $\Delta^2=M^2$  ($M$ is the nucleon mass)
through the combinations of nucleon DAs $V^p$, $A^p$ and $T^p$.

\end{itemize}

An elegant strategy allowing to ensure the polynomiality and the restricted
support properties for $\pi N$ TDAs consists in the use of the spectral representation
in terms of quadruple distributions \cite{Pire:2010if}. It generalizes for the TDA case
Radyushkin's double distribution representation for GPDs. The suggested approach of modelling
$\pi N$
TDAs is largely analogous to that employed for modelling nucleon
GPDs with the help of Radyushkin's double distribution Ansatz
\cite{RDDA4}.

However, contrary to GPDs, $\pi N$ TDAs lack a comprehensible
forward limit ($\xi=0$). Therefore, in order to work out the physical
normalization for  $\pi N$ TDAs it is illuminating to consider
the alternative limit $\xi=1$ in which  $\pi N$ TDAs are constrained by
 chiral dynamics and crossing due to the soft pion theorem.
A convenient form of the spectral representation for
$\pi N$ TDAs
reads   \cite{Lansberg:2011aa}:
\begin{eqnarray}
&&
H (w_i,\,v_i,\,\xi)=
\int_{-1}^1 d \kappa_i \int_{- \frac{1-\kappa_i}{2}}^{ \frac{1-\kappa_i}{2}} d\theta_i
\int_{-1}^1 d \mu_i \int_{- \frac{1-\mu_i}{2}}^{ \frac{1-\mu_i}{2}} d\lambda_i
\, \delta \big(w_i- \frac{\kappa_i-\mu_i}{2} (1-\xi) - \kappa_i \xi\big)  \nonumber \\ &&
\times
\delta\big(v_i- \frac{\theta_i-\lambda_i}{2} (1-\xi) - \theta_i \xi \big) \,
F(\kappa_i, \, \theta_i, \mu_i,\, \lambda_i ),
\label{Spectral_for_TDAs_redy_to_be_factorized}
\end{eqnarray}
where $F$ is a quadruple distribution.
The index $i=1,2,3$ here refers to one of three possible choices of independent variables
(quark-diquark coordinates):
$w_i= x_i-\xi$, $v_i= \frac{1}{2} \sum_{k,l=1}^3 \varepsilon_{ikl} x_k$.

We suggest to use the following factorized Ansatz for the quadruple distribution
$F$
in
(\ref{Spectral_for_TDAs_redy_to_be_factorized}):
\begin{eqnarray}
F(\kappa_i, \, \theta_i,\, \mu_i,\, \lambda_i )= 4 V(\kappa_i, \, \theta_i) \, h(\mu_i,\, \lambda_i),
\label{Factorized_ansatz}
\end{eqnarray}
where
$V(\kappa_i, \, \theta_i)$
is the combination of nucleon DAs
$V(y_1,y_2,y_3)$ ($\sum_{i=1}^3 y_i=1$)
to which
$\pi N$
TDA in question
reduces in the soft pion limit
$\xi=1$
expressed in terms of independent variables:
$\kappa_i= 2y_i-1$;
$\theta_i=  \sum_{k,l=1}^3 \varepsilon_{ikl} y_k$.

The profile function
$h(\mu_i,\, \lambda_i)$
is  normalized as
$
\int_{-1}^1 d \mu_i \int_{- \frac{1-\mu_i}{2}}^{ \frac{1-\mu_i}{2}} d\lambda_i \, h(\mu_i,\, \lambda_i) =1\,.
$
The support of the profile function
$h$ is also that of a baryon DA.
The simplest assumption for the profile is to take it to  be determined by the
asymptotic form of baryon DA
($120 y_1 y_2 y_3$
with
$\sum_{i=1}^3 y_i=1$)
rewritten in terms of variables
$\mu_i$, $\lambda_i$:
\begin{equation}
h(\mu_i,\, \lambda_i)=
\frac{15}{16} \, (1+\mu_i) ((1-\mu_i)^2-4 \lambda_i^2).
\label{Profile}
\end{equation}

Similarly to the GPD case
\cite{KSemenov_Polyakov:1999gs},
in order to satisfy the polynomiality condition in its complete form the spectral
representation for
$\pi N$
TDAs
$\{V_{1,2}, \,A_{1,2},\, T_{1,2} \}^{\pi N}$
should be supplemented with a $D$-term like contribution.
The simplest possible model for such a $D$-term
is the contribution of the $u$-channel nucleon exchange
into $\pi N$ TDAs computed in
\cite{Pire:2011xv}.
Thus we suggest a two component mode for
$\pi N$ TDAs which includes:
the spectral part  based on the factorized Ansatz for quadruple distributions
with input at $\xi=1$ from chiral dynamics;
and the nucleon exchange contribution as a $D$-term.
Phenomenological solutions for nucleon DAs (see {\it e.g.} \cite{Braun:2006hz})
are used as the numerical input for our model.

\section{Unpolarized cross section and single transverse target spin asymmetry}

In our factorized approach the leading order (both in
$\alpha_s$
and
$1/Q$)
amplitude of backward hard pion production
$\mathcal{M}^\lambda_{s_1s_2}$
reads \cite{Lansberg:2007ec}:
\begin{eqnarray}
{\cal M}^\lambda_{s_1s_2}={\cal{C}} \frac{1}{Q^4}
\bar U(p_2,s_2)
\Big[
\hat{\mathcal{E}}(\lambda)
 \gamma^5
{\cal{I} }(\xi,\Delta^2)+
\hat{\mathcal{E}}(\lambda)
\hat{\Delta}_T
 \gamma^5  {\cal{I}}'(\xi,\Delta^2) \Big] U(p_1,s_1).
\label{helicity_ampl_rewr}
\end{eqnarray}
We employ Dirac's hat notation $v_\mu \gamma^\mu \equiv \hat{v}$;
$\mathcal{E}$
denotes the polarization vector of the virtual photon and
$\bar{U}$, $U$
are the usual nucleon Dirac spinor.
$\mathcal{C}$ is the normalization constant
 $\mathcal{C} \equiv
-i
\frac{(4 \pi \alpha_s)^2 \sqrt{4 \pi \alpha_{em}} f_{N}^2}{ 54 f_{\pi} }$,
where
$\alpha_{em} (\alpha_s)$
stands for the electromagnetic (strong) coupling,
$f_\pi$
is the pion decay constant and
$f_N$
is the normalization constant of the nucleon wave function \cite{Chernyak_Nucleon_wave}.
The coefficients
$\cal{I}$, $\cal{I}'$
result from the calculation
of $21$ diagrams contributing to the hard scattering amplitude
(see \cite{Lansberg:2007ec}).

\begin{figure}
\begin{center}
\includegraphics[width=.45\textwidth]{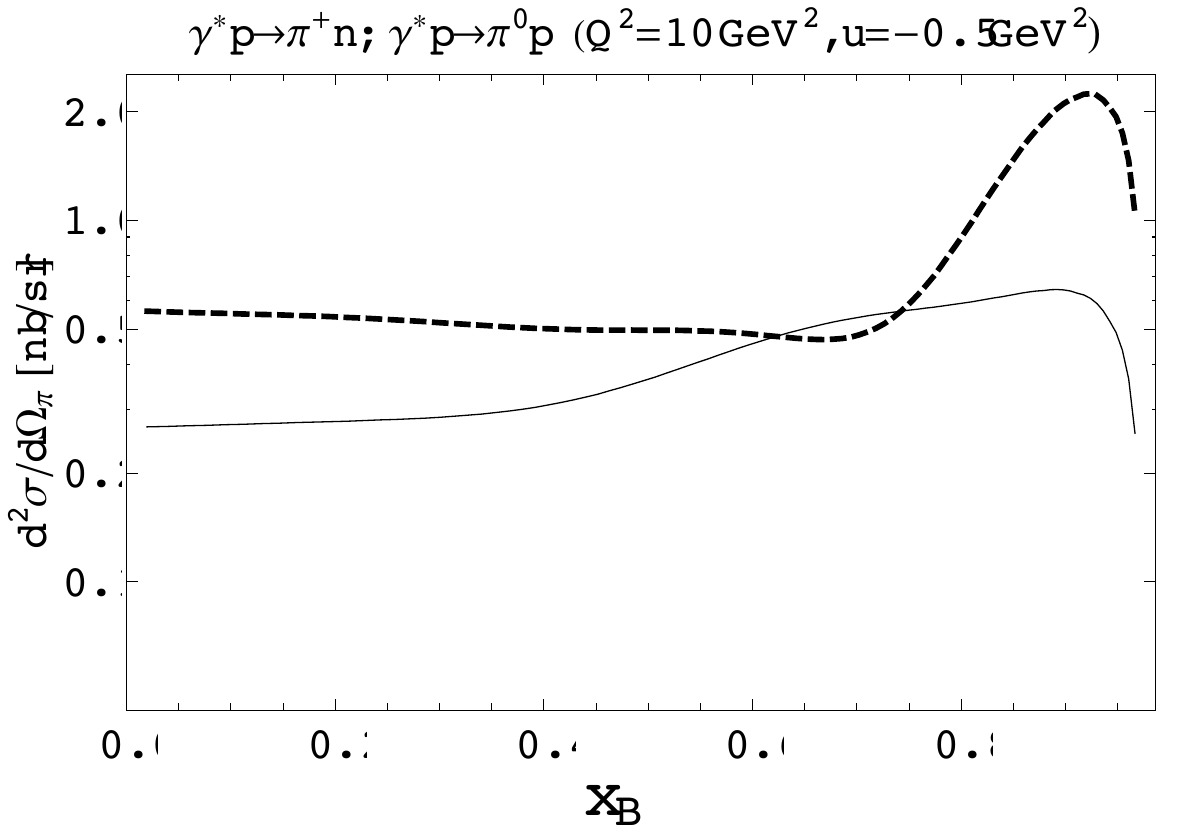}
\includegraphics[width=.45\textwidth]{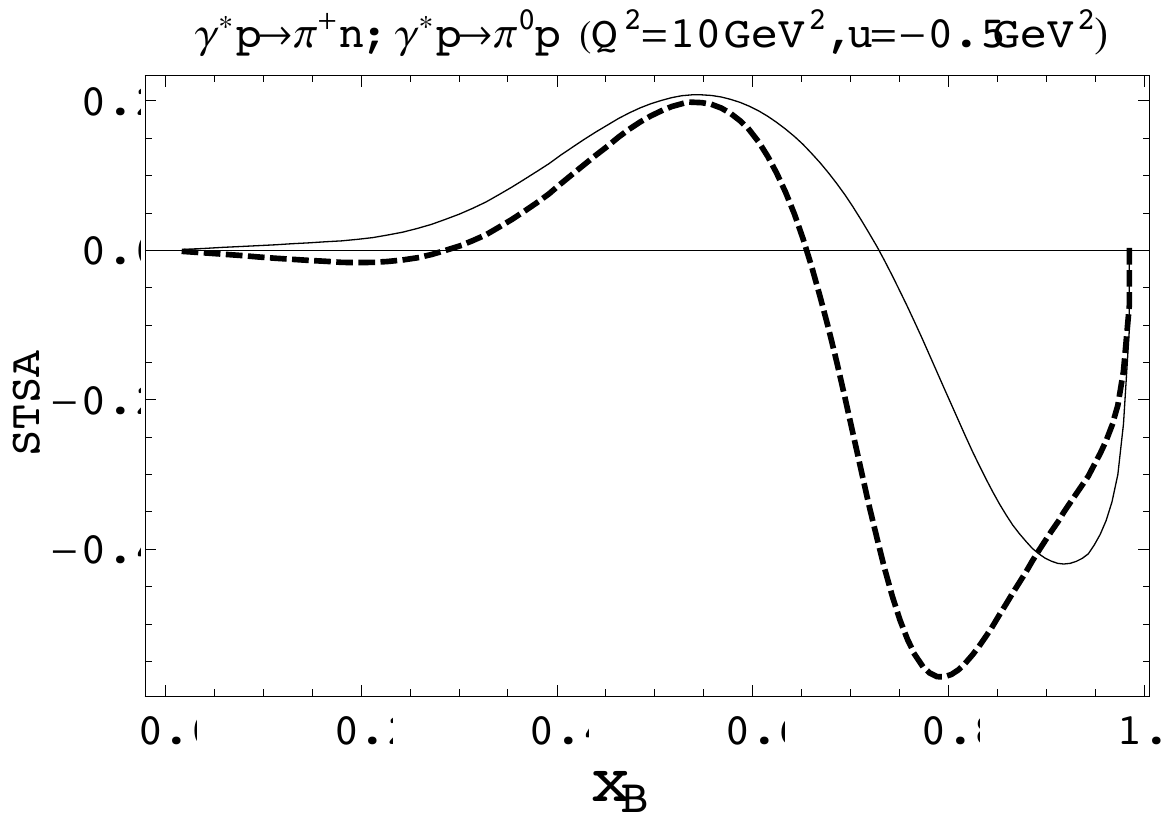}
\caption{
Unpolarized cross section ( {\bf left}) and single transverse target spin asymmetry ( {\bf right})
for backward $\pi^+$ (solid lines) and $\pi^0$ (dashed lines) production off proton.}
\label{fig2}
\end{center}
\end{figure}

Within the suggested factorization mechanism for backward pion electroproduction only the
transverse cross section
$\frac{d^2 \sigma_T}{d \Omega_\pi}$
receives a contribution at the leading twist level.
The unpolarized transverse cross section expresses as follows
through the coefficients
$\mathcal{I}$, $\mathcal{I}'$
introduced in
(\ref{helicity_ampl_rewr}):
\begin{eqnarray}
\frac{d^2 \sigma_T}{d \Omega_\pi}= |\mathcal{C}|^2 \frac{1}{Q^6}
\frac{\Lambda(s,m^2,M^2)}{128 \pi^2 s (s-M^2)} \frac{1+\xi}{\xi}
\big(
|\mathcal{I}|^2
-  \frac{\Delta_T^2}{M^2} |\mathcal{I}'|^2
\big),
\label{Work_fla_CS}
\end{eqnarray}
where
$
\Lambda(x,y,z)= \sqrt{x^2+y^2+z^2-2xy-2xz-2yz}
$
is the usual  Mandelstam function.
Within our two component model for $\pi N$ TDAs
$\mathcal{I}$ receives contributions both
from the spectral representation component and nucleon pole exchange contribution while
$\mathcal{I}'$ is determined solely by the nucleon pole contribution.
The scaling law for the unpolarized cross section (\ref{Work_fla_CS}) is
$1/Q^8$. This is to be compared with $1/Q^4$ behavior of the unpolarized cross section $\frac{d  \sigma }{d \Omega_\pi} $
in the conventional HMP regime.

On the left panel of Fig.~\ref{fig2}
we present our estimates for the unpolarized cross section
$\frac{d^2 \sigma_T}{d \Omega_\pi}$
of backward production of
$\pi^+$
and
$\pi^0$
off protons for
$Q^2=10 \, {\rm GeV}^2$
and $u=-0.5\, {\rm GeV}^2$
in ${\rm nb}/{\rm sr}$ as the function of $x_{B}$.
CZ solution \cite{Chernyak_Nucleon_wave}
for the nucleon DAs is used as phenomenological  input  for our model. The magnitude of
the  cross sections is large enough for a detailed investigation to be carried
at high luminosity experiments such as J-lab@6GeV  and especially J-lab@12GeV and EIC 
\cite{Boer:2011fh}. First
experimental results on backward pion electroproduction which would allow
to check validity of the  factorized description based on TDAs are expected to appear soon
\cite{Priv_J-lab}.

As a more sensitive observable to test the factorized description of hard reactions
it is convenient to consider asymmetries. These quantities, being the ratios of the cross
sections, are much less sensitive to the perturbative corrections.
For backward pion electroproduction an evident candidate is the single transverse target
spin asymmetry (STSA)
defined as:
\begin{eqnarray}
\mathcal{A}= \frac{1}{|\vec{s}_1|}
\frac{\left(
\int_0^\pi d \tilde{\varphi} |\mathcal{M}_{T}^{s_1}|^2 - \int_\pi^{2\pi} d \tilde{\varphi} |\mathcal{M}_{T}^{s_1}|^2
\right)}{ \left(
\int_0^{2\pi} d \tilde{\varphi} |\mathcal{M}_{T}^{s_1}|^2
\right) }
 = -\frac{4}{\pi} \frac{\frac{|\Delta_T|}{M}  \, {\rm Im} (\mathcal{I}'(\mathcal{I})^*)}{|\mathcal{I}|^2
- \frac{\Delta_T^2}{M^2} |\mathcal{I}'|^2}.
\label{Def_asymmetry}
\end{eqnarray}
Here $\tilde \varphi \equiv \varphi -\varphi_s$, where
$\varphi$ is the angle between leptonic and hadronic planes and $\varphi_s$ is the angle between
the leptonic plane and the transverse spin of the target $\vec{s}_1$.
On the right panel of Fig.~~\ref{fig2} we show the result of our calculation of the
STSA for backward
$\pi^+$
and
$\pi^0$
electroproduction off protons for
$Q^2=10 \, {\rm GeV}^2$ and $u=-0.5 \, {\rm GeV}^2$ as the function of $x_{B}$.
For backward pion electroproduction measurement of STSA, which according to our estimates
turns to be sizable in the valence region, should   be
considered as a crucial test of the applicability of our collinear factorized approach
for backward reactions.

This work is supported in part by the Polish NCN grant DEC-2011/01/B/ST2/03915
and by the French-Polish Collaboration Agreement Polonium.

\end{document}